# Simulation des Grandes Échelles d'un écoulement d'air turbulent pour le refroidissement d'amplificateurs de lasers

**M. BELLEC[a], U. BIEDER[b], N. LUCHIER[a], J.P. MORO[b], A. GIRARD[a], G. BALARAC[c]**

a. Univ. Grenoble Alpes, CEA, IRIG, DSBT, F-38000 Grenoble, morgane.bellec@cea.fr
b. CEA, Univ. Paris-Saclay, DEN-STMF, F-91191 Gif-sur-Yvette, ulrich.bieder@cea.fr
c. LEGI, CNRS-UGA-G-INP, F-38000 Grenoble, guillaume.balarac@grenoble-inp.fr

## Résumé :

*Le projet collaboratif français Trio4CLF a pour objectif de comprendre et de maitriser le refroidissement cryogénique d'amplificateurs de lasers à haute puissance (~1 PétaWatt) et à haut taux de répétition (1-10 Hertz). Dans de tels amplificateurs, le fluide évacue la puissance thermique déposée dans les plaques amplificatrices solides. Une connaissance fine de l'échange de chaleur et donc de l'écoulement turbulent est nécessaire pour évaluer l'impact du refroidissement sur la qualité du faisceau laser. Dans un premier temps, une Simulation des Grandes Échelles est conduite en air et sans chauffage afin d'étudier le développement de l'écoulement turbulent. Le code de CFD utilisé est TrioCFD, un code développé par le CEA. À des fins de validation, la simulation est menée dans la configuration de l'installation expérimentale : une soufflerie à boucle fermée appelée TRANSAT. Deux plaques séparées de 0,05 m sont placées horizontalement dans l'écoulement d'air de façon à représenter les plaques amplificatrices. Des couches limites turbulentes se développent à partir du bord d'attaque des plaques. Numériquement, l'écoulement d'entrée est un écoulement homogène plan de vitesse constante 10 m/s. Les résultats de cette Simulation des Grandes Échelles sont présentés ici sous la forme de l'étude du développement des couches limites turbulentes créées par les plaques.*

## Abstract:

*The French collaborative Trio4CLF project aims to understand and control the cryogenic cooling of amplifiers for high power (~1 PetaWatt) and high repetition rate (1-10 Hertz) lasers. In such amplifiers, the fluid evacuates the thermal power absorbed by the solid amplifying plates. A precise knowledge of the heat exchange and thus of the turbulent fluid flow in the amplifier is requested to evaluate its impact on the laser beam quality. As a first step, a Large Eddy Simulation is carried out in air without heating to study the development of the turbulent flow. The CFD code used is TrioCFD, a code developed by the CEA. For validation purpose, the simulation is carried out in the experimental setup configuration: a closed-loop wind tunnel called TRANSAT. Two horizontal plates, separated by 0.05 m, are put in the airflow to represent the amplifier plates. Turbulent boundary layers develop from the plates edges. Numerically, the entrance flow is a homogeneous planar flow with a constant velocity at 10 m/s. The results of this Large Eddy Simulation are presented in this paper as a study of the development of the turbulent boundary layers created by the plates.*

**Mots-clefs : Simulation des Grandes Échelles, Couche limite turbulence, Soufflerie, Laser haute puissance**

# 1 Introduction

La maitrise des aspects thermiques des amplificateurs solides est devenue un point-clef pour permettre l'augmentation des taux de répétition des lasers haute puissance [1]. Le Service des Basses Températures (SBT) du CEA Grenoble a proposé un système de refroidissement cryogénique multi-plaques en convection forcée [2] dans lequel le cristal amplificateur est refroidi par un écoulement turbulent d'hélium gazeux à basse température. Dans un tel système où le faisceau laser traverse le cristal et l'écoulement, comme schématisé Fig. 1, il est crucial de connaitre finement les variations de température. En effet, elles créent des variations d'indice optique qui risquent de nuire à la cohérence du faisceau. Il est pour cela nécessaire d'étudier finement l'écoulement turbulent.

Une littérature abondante est consacrée à l'étude numérique du développement de couche limite turbulente dans différentes conditions. On peut citer par exemple la Simulation Numérique Directe (SND) de Wu & Moin [3] d'une couche limite sur plaque lisse et sous gradient de pression nul. La couche y est déclenchée par des « blocs » de turbulence de grille introduits périodiquement depuis l'entrée de l'écoulement. Une méthode plus largement utilisée est celle développée par Lund *et al.* [4] consistant à recycler le champ de vitesse d'une position un peu à l'aval pour le réintroduire en entrée d'écoulement après l'avoir redimensionné. Schlatter & Örlü [5] soulignent l'importance des choix numériques et notamment du type de condition d'entrée en mettant en évidence les grands écarts entre les résultats de différents auteurs.

La stratégie adoptée par le projet Trio4CLF consiste à combiner des moyens expérimentaux et numériques de façon à pouvoir valider les choix numériques par comparaison aux mesures. On se place donc dans un premier temps dans le cas d'un écoulement d'air à température ambiante, dans des conditions proches de celles d'une soufflerie. Le cas d'étude et les moyens numériques employés sont résumés dans le deuxième paragraphe, suivis par la présentation dans le troisième paragraphe des premiers résultats numériques obtenus.

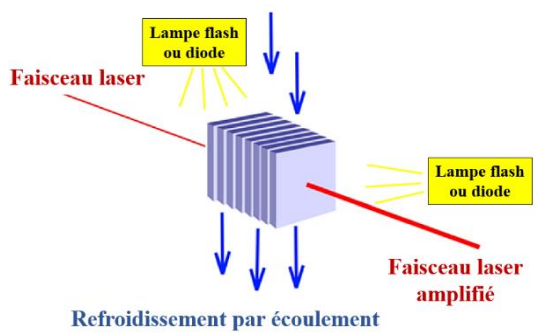
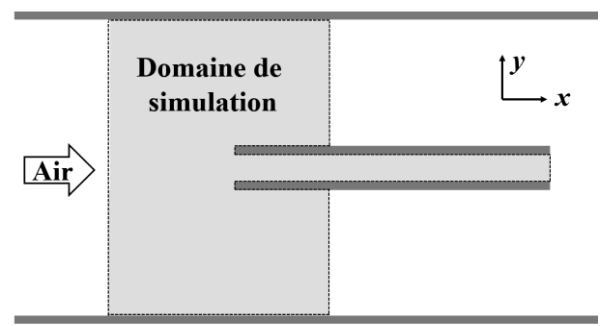

**Figure 1.** Vue schématique de l'amplificateur laser         **Figure 2.** Coupe transverse du domaine de calcul dans la veine de la soufflerie

# 2 Modèle numérique

L'écoulement qu'on souhaite simuler a les caractéristiques de la soufflerie TRANSAT dont une description détaillée est faite par Fougairolle [6]. Il s'agit d'une soufflerie en boucle fermée dont la veine d'essais est de section rectangulaire de largeur 0,6 m et de hauteur 0,5 m. L'écoulement entrant est de type turbulence de grille. La soufflerie est instrumentée par un système de mesures de vitesses et de températures par fils chauds et froids. On place à mi-hauteur de la veine d'essais deux plaques en couvrant la largeur complète, espacées de 0,05 m, dont la hauteur est 0,015 m et la longueur 0,5 m. Des nids d'abeille sont placés au-dessus et en dessous des plaques de façon à maintenir l'écoulement parallèle.

La simulation menée est de type Simulation des Grandes Échelles (SGE), ce qui consiste à résoudre les grandes échelles de la turbulence et à modéliser l'effet des petites échelles. La séparation est faite en filtrant les équations de Navier-Stokes. On pourra se référer à l'ouvrage de Sagaut [7] pour une description complète de la SGE et le détail des équations résolues. Le code utilisé est TrioCFD [8], un code de CFD open source développé par la Direction de l'Énergie Nucléaire du CEA-Saclay. Le modèle sous-maille adopté est le modèle WALE [9] ; aucune loi de paroi n'est utilisée. Le schéma temporel est un schéma Runge-Kutta du 3ème ordre. La discrétisation spatiale suit la méthode MAC [10] en utilisant des schémas de calcul centrés du 2ème ordre. On se place dans un premier temps dans le cas isotherme, et les propriétés de l'air sont prises constantes à température et pression atmosphériques.

Le domaine simulé est à trois dimensions ; un schéma en coupe selon la direction transverse (z) est présenté Fig. 2. Le domaine simulé commence 0,2 m avant le bord d'attaque des plaques et couvre la hauteur complète de la veine. La condition à la limite d'entrée est un profil de vitesse homogène de 10 m/s. Au-dessus et en dessous des plaques, une vitesse de sortie est imposée de façon à maintenir l'écoulement parallèle tandis qu'entre les plaques la condition de sortie est une pression homogène nulle. Le domaine de calcul a une profondeur de 0,15 m et est limité dans la direction transverse par une condition de périodicité. Le maillage est raffiné dans la zone d'intérêt entre les plaques. Dans cette zone, les mailles ont pour dimensions Δx=1,25 mm dans la direction longitudinale à l'écoulement, Δz=0,5 mm dans la direction transverse, et Δy variant dans la direction perpendiculaire à l'écoulement entre 0,08 mm près de la paroi et 0,6 mm au centre du canal. Au total, le domaine contient 27 millions de mailles.

## 3    Résultats numériques

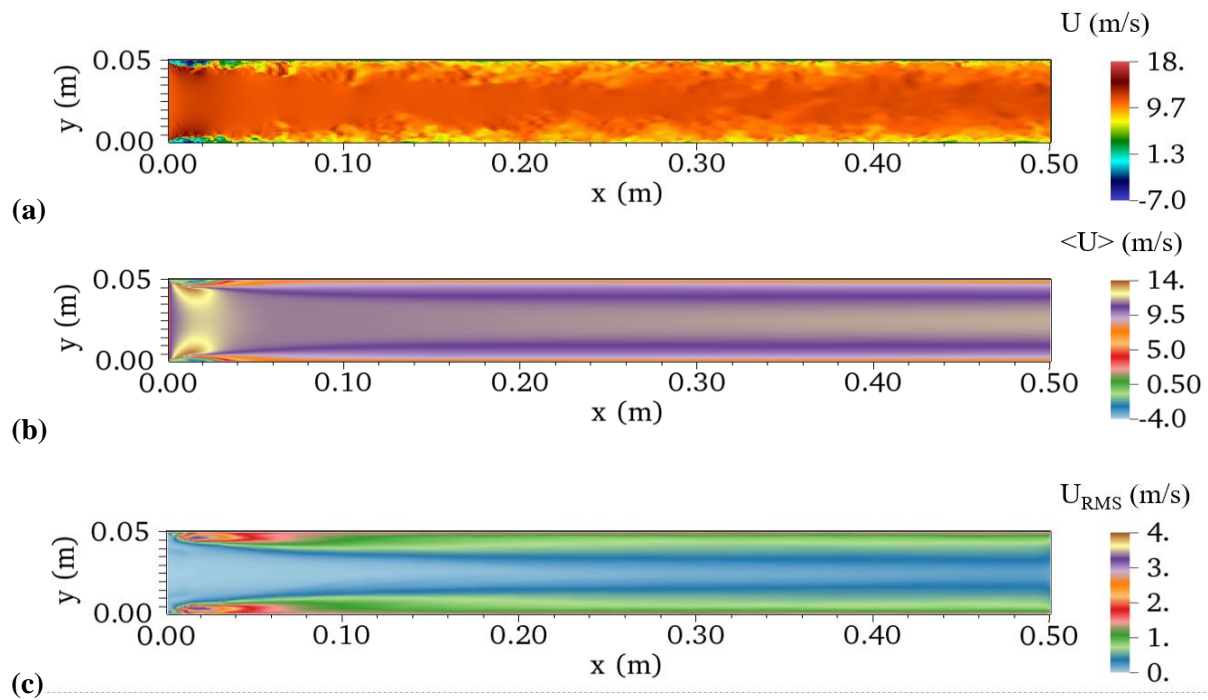

**Figure 3.** **(a)** Vitesse longitudinale instantanée U, **(b)** vitesse longitudinale moyenne <U> et **(c)** écart-type de vitesse longitudinale $U_{RMS}$ sur un plan de coupe (xy)

On présente Fig. 3 quelques résultats numériques obtenus dans la zone d'intérêt qui se situe entre les deux plaques sur une surface (xy) coupée selon la direction transverse d'homogénéité. Un champ de vitesse longitudinale instantanée U est tracé Fig. 3(a). L'écoulement d'air entrant, homogène à x=0 m, présente depuis le bord d'attaque un déclenchement de couche limite turbulente qui s'épaissit ensuite le long de l'écoulement. La Fig. 3(b) montre la vitesse longitudinale <U> moyennée pendant 0,65 secondes, soit 13 parcours moyens de canal. On y observe en début d'écoulement une zone de recirculation proche paroi où se créent des vitesses négatives compensées plus au centre par une zone opposée de fortes vitesses positives. La vitesse moyenne au centre du canal (à y=0,025 m) passe ensuite par un minimum à 11,2 m/s pour x=0,075 m avant d'augmenter régulièrement pour atteindre 11,6 m/s à la fin de l'écoulement. La couche limite turbulente est ainsi légèrement accélérée du fait de la présence de la seconde paroi. L'écart-type de la vitesse longitudinale $U_{RMS}$ est tracé Fig. 3(c). Son maximum est atteint au niveau du déclenchement de la couche limite, puis il s'amorti le long du canal. Cela peut être un effet de l'accélération [11]. Il faut toutefois souligner que l'écoulement amont n'est ici pas turbulent, et qu'il manque donc les mécanismes d'interaction de la couche limite avec l'écoulement extérieur.

## Conclusion

Une Simulation des Grandes Échelles d'un écoulement d'air a été menée dans le but d'étudier le refroidissement d'amplificateurs de laser haute puissance. Les plaques amplificatrices sont représentées en soufflerie par deux plaques lisses placées horizontalement dans l'écoulement parallèle d'air. Les premiers résultats numériques montrent qu'on se situe dans un cas d'écoulement complexe où le bord d'attaque des plaques déclenche le développement de deux couches limites turbulentes impactées par la présence d'une seconde plaque. Ces résultats devront être confrontés aux mesures expérimentales menées dans les mêmes conditions isothermes, puis avec un transfert de chaleur.

## Remerciements

Ces travaux ont bénéficié d'un accès aux moyens de calcul du CINES au travers de l'allocation de ressources 2019-A0052A10626 attribuée par GENCI.
Nous remercions pour son support financier le Programme Transverse de Compétences de simulation numérique du CEA, le Commissariat à l'Énergie Atomique et aux Énergies Alternatives.

## Références

[1] B. Le Garrec, Laser-diode and Flash Lamp Pumped Solid-State Lasers, in: AIP Conf. Proc., AIP, 2010: pp. 111–116.
[2] J.P. Perin, F. Millet, B. Rus, M. Divoký, Cryogenic Cooling For High Power Laser Amplifiers, in: Singapore, 2011.
[3] X. Wu, P. Moin, Direct numerical simulation of turbulence in a nominally zero-pressure-gradient flat-plate boundary layer, J. Fluid Mech. 630 (2009) 5.
[4] T.S. Lund, Generation of Turbulent Inflow Data for Spatially-Developing Boundary Layer Simulations, in: 1996.
[5] P. Schlatter, R. öRlü, Assessment of direct numerical simulation data of turbulent boundary layers, J. Fluid Mech. 659 (2010) 116–126.
[6] P. Fougairolle, Caractérisation expérimentale thermo-aéraulique d'un jet transverse impactant ou non, en turbulence de conduite, PhD Thesis, Université Joseph-Fourier-Grenoble I, 2009.
[7] P. Sagaut, Large Eddy Simulation for Incompressible Flows, Springer Berlin Heidelberg, Berlin, Heidelberg, 2002.
[8] P.-E. Angeli, U. Bieder, G. Fauchet, Overview of the TrioCFD code: main features, V&V procedures and typical applications to nuclear engineering, in: Proc. 16th Int. Top. Meet. Nucl. React. Therm. Hydraul. NURETH-16, 2015.
[9] F. Nicoud, F. Ducros, Subgrid-scale stress modelling based on the square of the velocity gradient tensor, Flow Turbul. Combust. 62 (1999) 183–200.
[10] F.H. Harlow, J.E. Welch, Numerical Calculation of Time-Dependent Viscous Incompressible Flow of Fluid with Free Surface, Phys. Fluids. 8 (1965) 2182.
[11] U. Piomelli, J. Yuan, Numerical simulations of spatially developing, accelerating boundary layers, Phys. Fluids. 25 (2013) 101304.